\documentclass[twocolumn,pra,aps,superscriptaddress,10pt]{revtex4}
\usepackage{MnSymbol}
\usepackage{amsmath,amsfonts}
\usepackage{graphicx,xcolor}
\usepackage[english]{babel}
\usepackage{mathrsfs}
\usepackage{ytableau}
\usepackage[bookmarks=true, color links, linkcolor=blue, urlcolor=blue, citecolor=blue, plainpages=false, pdfpagelabels, final, breaklinks=true]{hyperref}
\begin{document}

\newcommand\bra[1]{\mathinner{\langle{\textstyle#1}\rvert}}
\newcommand\bbra[1]{\mathinner{\llangle{\textstyle#1}\rvert}}
\newcommand\ket[1]{\mathinner{\lvert{\textstyle#1}\rangle}}
\newcommand\kket[1]{\mathinner{\lvert{\textstyle#1}\rrangle}}
\newcommand\braket[1]{\mathinner{\langle{\textstyle#1}\rangle}}
\newcommand\ketbra[2]{\mathinner{\lvert{\textstyle#1}\rangle\langle{\textstyle#2}\rvert}}
\newcommand\aavg[1]{\mathinner{\left\llangle{\textstyle#1}\right\rrangle}}
\newcommand\hata{\hat{a}}
\newcommand\hatb{\hat{b}}
\newcommand\hatc{\hat{c}}
\newcommand\hatH{\hat{H}}
\newcommand\hatN{\hat{N}}
\newcommand\hatU{\hat{U}}
\newcommand\hatV{\hat{V}}
\newcommand\hatW{\hat{W}}
\newcommand\calA{\mathcal{A}}
\newcommand\calC{\mathcal{C}}
\newcommand\calD{\mathcal{D}}
\newcommand\calH{\mathcal{H}}
\newcommand\calL{\mathcal{L}}
\newcommand\calS{\mathcal{S}}
\newcommand\calU{\mathcal{U}}
\newcommand\calY{\mathcal{Y}}
\newcommand\calZ{\mathcal{Z}}
\newcommand{\varF}{\mathscr{F}}
\newcommand{\varL}{\mathscr{L}}
\newcommand{\varM}{\mathscr{M}}
\newcommand\msc[1]{\textbf{\textcolor{red}{MSC: #1}}}


\title{Symmetry and Liouville Space Formulation of Decoherence-Free Subsystems}

\author{Mi-Jung So}
\affiliation{Department of Physics, Korea University, Seoul 02841, South Korea}
\affiliation{School of Quantum, Korea University, Seoul 02841, South Korea}


\author{Mahn-Soo Choi}
\email{choims@korea.ac.kr}
\affiliation{Department of Physics, Korea University, Seoul 02841, South Korea}
\affiliation{School of Quantum, Korea University, Seoul 02841, South Korea}

\begin{abstract}
we propose a generic and systematic decoherence-free scheme to encode quantum information into an open quantum system based focusing on symmetry. Under a given symmetry, the Liouville space is decomposed into invariant subspaces characterized by a tensor-product structure. A decoherence-free subsystem is then identified as a factor of the tensor product.
Unlike decoherence-free subspaces, which typically require strong symmetries, decoherence-free systems are permitted under less restrictive weak symmetries. Specifically, we primarily concern the permutation symmetry in conjunction with the unitary symmetry and utilize the Schur-Weyl duality, which facilitates numerous efficient and systematic calculations based on the well-established group representation theory. Employing the isomorphism between the Liouville space and the fictitious Hilbert space, we construct a super-Schur basis, which block-diagonalizes the super-operators that describe the noisy quantum channels, both in the Kraus representation and in terms of the quantum master equation. Each block reveals the tensor-product structure and facilitates the identification of physically relevant decoherence-free subsystems under the specified weak symmetry.
\end{abstract}

\maketitle

\section{Introduction}

Protecting quantum information from environmental quantum noise remains one of the central challenges in the development of robust, scalable quantum technologies. In realistic scenarios, a quantum system inevitably interacts with its surrounding environment, leading to quantum decoherence, the loss of quantum coherence as the system's information becomes entangled with uncontrollable external degrees of freedom \cite{Zurek03a,Palma96a}. This general phenomenon not only degrades the performance of quantum computing and communication protocols, but can also fundamentally limit the advantages offered by quantum coherence and entanglement .

A major approach to combat decoherence is quantum error correction, which actively monitors and corrects errors in quantum states using encoded logical qubits and a sequence of high-precision measurements and gate operations. Quantum error correction has achieved numerous theoretical and experimental milestones, such as the development of fault-tolerant thresholds, stabilizer and topological codes, and successful implementation in trapped ions and superconducting platforms \cite{Shor95a,Gottesman96a,Gottesman98a,Kitaev03a,Postler22a,GoogleQuantumAI23a,Bluvstein23a,Acharya24a,Aghaee25a}. However, these active protocols incur significant resource overhead, demanding both a large number of physical qubits and highly accurate operations consistently across time and hardware.

In contrast, a passive approach to decoherence control leverages structural properties of system-environment interactions to encode quantum information in such a way that it inherently remains unaffected by specific types of environmental noise. This idea was formalized through the development of decoherence-free subspaces, subspaces of the system’s Hilbert space that are invariant under the action of certain noise operators \cite{Lidar98a,Zanardi97a,Zanardi97b,Duan98a}. When the system-environment coupling exhibits symmetry, such as collective decoherence, quantum information encoded in a DFS can evolve unitarily despite the presence of environmental interactions.

The decoherence-free subspace framework was later generalized to the more encompassing notion of decoherence-free subsystems (also known as noiseless subsystems), a concept grounded in operator algebra and representation theory \cite{Knill00a,Zanardi00a,bacon2000universal,Lidar03a,Shabani05a}. Rather than requiring an invariant subspace, this approach allows for the identification of subsystems that remain dynamically isolated from the noise under general symmetric couplings. This generalization has led to a powerful and unified formalism for identifying noise-protected encodings based on the structure of the noise algebra.

Despite these theoretical advances, a key practical challenge remains unsolved: for a given noise model or symmetry present in the quantum evolution, no general or systematic method exists for constructing the relevant decoherence-free subspaces/subsystems. Most existing approaches rely heavily on case-specific insight or computationally intensive algebraic analysis of the Lindblad or Kraus operators \cite{Wang13a,Viola00a,Barenco97a,singh:2402.18703}. Furthermore, in the Liouville-space or superoperator formalism, where quantum operations are treated as linear maps on operator space, the challenge of identifying symmetry-protected subsystems becomes more intricate and has not been systematically addressed.

There is thus a compelling need for a systematic, symmetry-exploiting approach to construct decoherence-free structures in Liouville space. Such a framework would not only deepen our theoretical understanding of decoherence and open system dynamics but also accelerate the development of practical encoding and control strategies in near-term quantum devices, where full-scale QEC may remain out of reach.

In this work, we propose a comprehensive and systematic methodology for encoding quantum information into open quantum systems by exploiting system symmetries. 
For a given symmetry, we partition the Liouville space into invariant subspaces exhibiting a tensor-product structure. Subsequently, decoherence-free subsystems are identified as specific factors within these tensor products.
In contrast to decoherence-free subspaces, which typically necessitate stringent symmetry requirements, our approach facilitates the existence of decoherence-free subsystems under more relaxed, weak symmetry conditions. 
Specifically, we concentrate on the interplay between permutation symmetry and unitary symmetry, employing Schur-Weyl duality to facilitate efficient and rigorous analysis rooted in group representation theory. 
By leveraging the isomorphism between the Liouville space and a corresponding fictitious Hilbert space, we construct a super-Schur basis that block-diagonalizes the pertinent super-operators governing noisy quantum channels, both in the Kraus representation and in the form of quantum master equations. 
This block-diagonalization explicitly unveils the underlying tensor-product structure and facilitates the streamlined identification of decoherence-free subsystems supported by the weak symmetry constraints.

This article is organized as follows: 
Section~\ref{schur-basis-for-liouville-space} discusses the decomposition of the Liouville space into invariant subspaces under the permutation symmetry, examines the tensor-product structure of each subspace, and constructs the corresponding super-Schur basis that block-diagonalizes the quantum operations.
Sections~\ref{sec:symmetry-kraus} and \ref{sec:symmetry-lindblad} provide the general notions of both strong and weak super-symmetries in noisy quantum channels formulated in the Liouville space. 
Section~\ref{sec:symmetry-kraus} presents detailed examples of permutation-symmetric noisy quantum channels in the Kraus representation and demonstrates how the super-Schur basis facilitates the identification of decoherence-free subsystems in super-operators. 
Section~\ref{sec:symmetry-lindblad} provides examples of permutation-symmetric Lindblad equations and explains how the super-Schur basis facilitates identifying decoherence-free subsystems efficiently. 
Finally, Section~\ref{conclusion} concludes the article. 
To deliver the points clearer, we put some technical details in appendicess focusing on the major points in the main text:
Appendix~\ref{Appendix:A} provides details about block-diagonalization
of the super-operators. Appendix~\ref{Appendix:B} demonstrates how to calculate the number of the irreducible representations.

\section{Schur Basis for Liouville space}
\label{schur-basis-for-liouville-space}

\def\Usch{U_{\text{Sch}}}
\def\cN{{\cal N}}
\def\cI{{\cal I}}
\def\cP{{\cal P}}
\def\cQ{{\cal Q}}		 
\def\bp{{\bf p}}
\def\bq{{\bf q}}
\newcommand{\oprod}[1]{\l| #1 \r\rangle\!\!\l\langle #1 \r|}


This work aims to propose a generic symmetry-based decoherence-free way to encode quantum information into a system of multiple \emph{qudits} that are subject to quantum noise. To do that, in this section we first examine the structure of the Liouville space associated with the system, especially, by decomposing it into subspaces that are not only invariant under a given set of symmetry transformations but also feature a tensor-product structure.
This enables to identify a relevant decoherence-free subsystem under the symmetry in question with a certain factor of the tensor products.
Note that we deliberately distinguish the notion of decoherence-free \emph{subsystem} under the so-called \emph{weak} symmetries \cite{Knill00a,Zanardi00a,DeFilippo00a,Kempe01a} from the decoherence-free \emph{subspace} under the far more restrictive \emph{strong} symmetries \cite{Zanardi97a,Zanardi97b,albertSymmetriesConservedQuantities2014}.

In this work, we will primarily focus on the permutation symmetry in conjunction with the unitary symmetry: The permutation symmetry prevails in most physical situations, especially with independ and identically distributed (i.i.d.) states and channels, and is definitely a top priority in symmetry-related investigations.
On the other hand, the unitary symmetry is expected in more restricted circumstances. Nevertheless, we will regard the unitary symmetry almost on the equal footing as the permutation symmetry because these two symmetries are intricately connected with each other as summarized in the celebrated Schur-Weyl duality \cite{Fulton04a,Goodman09a}.
One representative aspect of the Schur-Weyl duality is marked in the fact that the actions of the permutation and unitary symmetries commute with each other.
Thanks to this duality, many details about the irreducible representations of the both symmetries are well-known and can be calculated systematically,\cite{Fulton04a,Goodman09a} allowing us to provide explicit examples of our framework.
The Schur-Weyl daulity has found many other applicaitons in a wide range of areas such as quantum information theory, quantum algorithms, and many-body physics.

While the permutation symmetry is our main concern and the unitary symmetry plays an auxiliary role, our framework can be directly switched to the case with the unitary symmetry as per the Schur-Weyl duality. It can be easily extended to any other symmetry as well, so long as the group-theoretic structure associated with the symmetry is well established.

\subsection{Schur-Weyl Duality in Liouville space}
\label{schur-weyl-duality-in-liouville-space}

We start by extending the features of the Schur-Weyl duality on a Hilbert space (for pure-state vectors) to a Liouville space (for mixed-state density matrices), and later construct an efficient basis for the decoherence-free subsystems. 

Consider a quantum system comprising \(n\) qudits associated with the tensor-product Hilbert space $\mathcal{H}_{d}^{\otimes n}$ with each qudit associated with a Hilbert space \(\mathcal{H}_{d}\) of dimension \(d\).
The Liouville space 
\begin{math}
\mathcal{L}(\mathcal{H}_{d}^{\otimes n})
\end{math}
is the vector space of all linear operators on \(\mathcal{H}_d^{\otimes n}\), and has dimension \(d^{2n}\) in accordance with the isomorphic relation
\begin{math}
\mathcal{L}(\mathcal{H}_{d}^{\otimes n})
\equiv \mathcal{L}(\mathcal{H}_d)^{\otimes d}
\cong\mathcal{H}_{d^{2}}^{\otimes n}
\end{math}
\cite{endnote:ChoiIsomorphism}.
It is common to equip the Liouville space with a Hermitian product (often called the Hilbert-Schmidt inner product) defined by 
\begin{equation}
\label{Paper::eq:HSinner}
\llangle \hat{A}, \hat{B} \rrangle
:= d^{-n}\mathrm{Tr}[\hat{A}^{\dagger} \hat{B}] ,
\end{equation}
where factor $d^{-n}$ accounts for the normalization convention in this work.

The Liouville space grows exponentially with the number $n$ of qudits, so it is impractical to handle it directly.
For practical calculations, it is essential to decompose it into smaller subspaces invariant under known symmetries.
To do it efficiently, having the permutation symmetry in mind, we adopt the idea employing the Schur-Weyl duality, originally put forward on the usual tensor-product Hilbert space \cite{Fulton04a,Goodman09a}.
We decompose the Liouville space (rather than the Hilbert space)
into irreducible Liouville subspaces
invariant under the symmetric group \(\mathcal{S}_n\) and the unitary group
\(\mathcal{U}_{d^{2}}\) \cite{endnote:1}
as
\begin{equation}
\label{eq:duality}
\mathcal{L}(\mathcal{H}_d^{\otimes n}) 
\simeq \mathcal{H}_{d^{2}}^{\otimes n} 
= \sum_{\lambda}\mathcal{Y}^{\lambda}\otimes \mathcal{W}^{\lambda},
\end{equation}
where \(\mathcal{Y}^\lambda\) and \(\mathcal{W}^{\lambda}\) denote  
irreducible representation spaces
of \(\mathcal{S}_{n}\) and \(\mathcal{U}_{d^{2}}\), respectively.
Here, each component is labeled systematically by a partition
\begin{math}
\lambda := (\lambda_{1}, \lambda_{2}, \ldots, \lambda_{k})
\end{math}
of  integer \(n\) into $k$ integers,
\begin{math}
n = \lambda_1 + \lambda_2 + \cdots + \lambda_k,
\end{math}
with 
\(\lambda_{1} \geq \lambda_{2} \geq \ldots \geq \lambda_{k} \geq 0\)
and
$k\leq\min(n,d^2)$.
Typically, a partition $\lambda$ is depicted by a Young diagram of $n$ boxes arranged in $k$ rows (in this article, we will use the partition and Young diagram interchangeably).
For example, partition $\{4,2,1\}$ of 7 is represented by Young diagram
\begin{equation}
\label{eq:ydiagram}
\ytableausetup{smalltableaux,aligntableaux=bottom}
\ydiagram{4,2,1} . 
\end{equation}
Naturally, just like for the tensor-product Hilbert space,
the total number of inequivalent irreducible Liouville subspaces 
that are invariant under $\mathcal{S}_n$ or $\mathcal{U}_{d^2}$
is determined by the number of partitions $\lambda$ of \(n\) with $d^2$ rows; see Appendix \ref{Appendix:B} for details. 

Note that the tensor-product form of each component in \eqref{eq:duality} implies the existence of decoherence-free subsystems in the sense of strong and weak symmetries, as we will see in Section~\ref{sec:decoherence-free-subsystem}.

Note also that while any symmetry group enables a decomposition of the Liouville space similar to Eq.~\eqref{eq:duality}, there are not so many known systematic (not to mention efficient) ways to calculate the decomposition for general symmetry groups.
On the other hand, for the permutation and/or unitary symmetries,
the Schur-Weyl duality provides a systematic way called the Schur transform \cite{Fulton04a,Goodman09a} as explained below. 
Interestingly, the Schur transform, the central technical part of the Schur-Weyl duality, is known to be efficiently calculated on a quantum computer.\cite{baconEfficientQuantumCircuits2006,Kirby18a,Krovi19a}

\subsection{Super-Schur Basis}
\label{sec:super-schur-basis}

The Schur transform is a linear map from the standard tensor-product basis to the so-called Schur basis that transforms within the irreducible subspaces in Eq.~\eqref{eq:duality}. 
For most physical purposes, the construction of such a basis is a concrete and efficient way for actutual decomposition of the Liouville space as in Eq.~\eqref{eq:duality}.

Here, instead of developing the Schur transform from the first principles, we exploit the isomorphism
\begin{math}
\mathcal{L}(\mathcal{H}_{d}^{\otimes n}) \simeq
\mathcal{H}_{d^2}^{\otimes n},
\end{math}
and directly maps the Schur basis for the Hilbert space 
\begin{math}
\mathcal{H}_{d^2}^{\otimes n}
\end{math}
to the Liouville space
\begin{math}
\mathcal{L}(\mathcal{H}_{d}^{\otimes n})
\end{math}
\cite{endnote:ChoiIsomorphism}.
To understand the resulting basis, which we call the \emph{super-Schur basis} to distinguish it from the Schur basis for Hilbert spaces, note that the basis elements $|Y^{\lambda}\rrangle$ for $\mathcal{Y}^\lambda$ are associated with the standard Young tableau $Y^{\lambda}$ of shape $\lambda$ \cite{Goodman09a,Sagan01a}.
Similarly, the basis elements $|W^\lambda\rrangle$ for $\mathcal{W}^\lambda$ may be specified by the Weyl tableaux (or semi-standard Young tableaux) $W^\lambda$ of shape $\lambda$ of degree $d^2$ \cite{Goodman09a,Sagan01a}.
Combining these two bases results in the super-Schur basis
\begin{math}
\{ |Y^{\lambda},W^{\lambda}\rrangle \}
\end{math}
labeled by the standard Young tableaux $Y^\lambda$ of shape $\lambda$, the Weyl tableaux $W^\lambda$ of shape $\lambda$ and degree $d^2$, and the partition $\lambda$ of integer $n$ \cite{baconEfficientQuantumCircuits2006,Krovi19a,Kirby18a}.
As such, the dimensions of subspaces $\mathcal{Y}^\lambda$ and $\mathcal{W}^\lambda$ are given by the number of standard Young and Weyl tableaux, respecitvely. These dimensions determine the capacity of decoherence-free encoding of quantum information.
Recall that the numbers of standard Young and Weyl tableaux for a given partition \(\lambda\) can be easily computed using well-known formulae \cite{Sagan01a}. 

The actual expression of each basis element $|Y^\lambda,W^\lambda\rrangle$ in terms of tensor-product basis elements is determined by the Schur transform as detailed in Refs.~\cite{baconEfficientQuantumCircuits2006,Krovi19a,Kirby18a}.
For example, in a system of three qubits ($n=3$ and $d=2$),
\begin{equation}
\ytableausetup{smalltableaux,centertableaux}
\left|\,
\begin{ytableau} 1 & 2\\ 3\end{ytableau},
\begin{ytableau} 2 & 2\\ 3\end{ytableau}\,
\right\rrangle
= \sqrt{\frac{2}{3}}\hat{X}\otimes\hat{X}\otimes\hat{Y}
- \frac{1}{\sqrt{6}}\hat{X}\otimes\hat{Y}\otimes\hat{X}
- \frac{1}{\sqrt{6}}\hat{Y}\otimes\hat{X}\otimes\hat{X},
\end{equation}
which corresponds to
\begin{equation}
\ytableausetup{smalltableaux,centertableaux}
\left|\,
\begin{ytableau} 1 & 2\\ 3\end{ytableau},
\begin{ytableau} 2 & 2\\ 3\end{ytableau}\,
\right\rangle
= \sqrt{\frac{2}{3}}\ket{1,1,2}
- \frac{1}{\sqrt{6}}\ket{1,2,1}
- \frac{1}{\sqrt{6}}\ket{2,1,1}
\end{equation}
under the correspondences 
\begin{math}
\ket{0} \leftrightarrow \hat{I},
\end{math}
\begin{math}
\ket{1} \leftrightarrow \hat{X},
\end{math}
\begin{math}
\ket{2} \leftrightarrow \hat{Y},
\end{math}
and
\begin{math}
\ket{3} \leftrightarrow \hat{Z}
\end{math}
(recall the normalization convention in Eq.~\eqref{Paper::eq:HSinner})
for the isomorphism
\begin{math}
\mathcal{L}(\mathcal{H}_{d}^{\otimes n}) \simeq
\mathcal{H}_{d^2}^{\otimes n}
\end{math}
\cite{endnote:ChoiIsomorphism}.
Here, $\hat{I}$ is the identity matrix, and $\hat{X}$, $\hat{Y}$ and $\hat{Z}$ are the Pauli operators on a single qubit.

\section{Symmetry in Kraus Representation}
\label{sec:symmetry-kraus}

A quantum decoherence or noisy quantum channel that corrupts quantum states of an open system is described by a trace-preserving completely positive linear map,
\begin{math}
\hat\rho \mapsto \mathscr{F}(\hat\rho),
\end{math}
on density operator $\hat\rho$.
Such a map, which hereafter we call a \emph{super-operator} for convenience,
may be put in the Kraus representation
\begin{equation}
\mathscr{F} (\hat{\rho})=\sum_{\mu}\hat{F}_{\mu}\hat{\rho}\hat{F}_{\mu}^{\dagger},
\end{equation}
where the Kraus operators $\hat{F}_{\mu}$ associated with $\mathscr{F}$ are responsible for different quantum decoherence processes labeled by index $\mu$.
The Kraus operators satisfy the closure relation $\sum_{\mu}\hat{F}_{\mu}^{\dagger}\hat{F}_{\mu}=\hat{I}$, where $\hat{I}$ is the identity operator, which ensures that the noisy quantum channel is trace-preserving.
Throughout this work, we assume without loss of generality that the Kraus operators $\hat{F}_\mu$ are mutually orthogonal with respect to the Hilbert-Schmidt inner product in Eq.~\eqref{Paper::eq:HSinner}.

In this section, we examine the symmetry properties of quantum channels in terms of the Kraus representation.
As mentioned before, in this work, we will focus on the permutation symmetry, which is described by the symmetric group $\mathcal{S}_n$.
However, all arguments below hold for any symmetry described by a finite group.

\subsection{Permutation Symmetry in Kraus Representation}
\label{sec:perm-symmetry-kraus}

On the Hilbert space \(\mathcal{H}_d^{\otimes n}\), a permutation $\pi$ in the symmetry group $\mathcal{S}_n$ is represented by operator $\hat\pi$ defined as
\begin{equation}
\hat\pi\ket{x_1x_2\cdots x_n} := 
\ket{x_{\pi^{-1}_1}x_{\pi^{-1}_2}\cdots x_{\pi^{-1}_n}}
\end{equation}
for a standard tensor-product basis state $\ket{x_1x_2\cdots x_n}$ in \(\mathcal{H}_d^{\otimes n}\).
That is, the set of permutation operators,
\begin{math}
\left\{ \hat{\pi} \,|\, \pi\in\calS_n \right\},
\end{math}
acting on the operators of \(\mathcal{H}_d^{\otimes n}\) serves as the representation
of \(\mathcal{S}_{n}\).
Similarly, on the Liouville space \(\mathcal{L}(\mathcal{H}_{d}^{\otimes n})\),
permutation $\pi$ is represented by super-operator $\mathscr{S}_\pi$ defined by
\begin{equation}
\mathscr{S}_{\pi} (\hat{\rho}) := \hat{\pi} \; \hat{\rho} \; \hat{\pi}^{\dagger}
\end{equation}
for a density operator $\hat\rho$ in $\mathcal{L}(\mathcal{H}_d^{\otimes n})$.
The set of permutation super-operators
\(\left\{ \mathscr{S}_{\pi}\, |\, \hat{\pi}\in\calS_n \right\}\)
forms another representation of the symmetric group. 

A permutation-symmetric quantum channel can be described by super-operators $\mathscr{F}$ that commute with all permutation super-operators;
\begin{equation}
\label{Paper::eq:cmmSuperOp}
\mathscr{S}_{\pi} \mathscr{F} = \mathscr{F} \mathscr{S}_{\pi}
\end{equation}
for all permutations $\pi$ in $\calS_n$.
In terms of the Kraus representation, this is equivalent to the requirement that
\begin{equation}
\sum_{\mu} \mathscr{S}_{\pi}(\hat{F}_{\mu}) \hat{\rho} \mathscr{S}_{\pi} (\hat{F}_{\mu}^{\dagger})
= \sum_{\mu} \hat{F}_{\mu} \hat{\rho} \hat{F}_{\mu}^{\dagger}
\end{equation}
for any operator $\hat\rho$.
From the unitary freedom of the Kraus operators~\cite{endnote:2},
this implies that for every permutation $\pi$,
there exists a unitary matrix $U(\pi)$ such that
\begin{equation}
\label{Paper::eq:KrausUnitaryFreedom}
\hat\pi\,\hat{F}_{\nu}\hat\pi^\dag
= \sum_{\mu}\hat{F}_{\mu}U(\pi)_{\mu \nu}
\end{equation}
for all Kraus operators $\hat{F}_\nu$.
In turn, this means that the set of Kraus operators $\{\hat{F}_\mu\}$ associated with the permutation-symmetric super-operator $\varF$ spans a representation space (which is a subspace in the Liouville space)  of the symmetric group $\calS_n$, and $U$ is the unitary representation of $\calS_n$ on this space.

Of particular interest is the case of \emph{local} quantum channels:
In this case, the Kraus operators are given by tensor-products
\begin{math}
\hat{F}_\mu = \hat{f}_{\mu_1}\otimes \cdots\otimes\hat{f}_{\mu_n}
\end{math}
of single-qudit (or single-particle) Kraus operators $\hat{f}_{\mu_k}$ of type $\mu_k$ on qudit (or particle) $k$; the index $\mu$ on the left-hand side is then regarded as a collective index
\begin{math}
\mu := (\mu_1,\cdots,\mu_n).
\end{math}
The action of permutation $\pi$ on such a Kraus operator is simply given by the permutation of single-qudit Kraus operators,
\begin{equation}
\hat\pi \left(\hat{f}_{\mu_1}\otimes \cdots\otimes\hat{f}_{\mu_n}\right) \hat\pi^\dag
= \hat{f}_{\mu_{\pi^{-1}_1}}\otimes \cdots\otimes\hat{f}_{\mu_{\pi^{-1}_n}}.
\end{equation}
Therefore, any permutation-symmetric local super-operator 
is represented in the form 
\begin{multline}
\mathscr{F}(\hat\rho) =
\sum_{\mu_1\cdots\mu_n}
\sum_{\pi\in\mathcal{S}_n}
\left(\hat{f}_{\mu_{\pi^{-1}_1}}\otimes\cdots\otimes
\hat{f}_{\mu_{\pi^{-1}_n}}\right)
\hat\rho \\{}\times
\left(\hat{f}_{\mu_{\pi^{-1}_1}}\otimes\cdots\otimes
\hat{f}_{\mu_{\pi^{-1}_n}}\right)^\dag .
\end{multline}
Physically, local quantum channels are relevant when decoherence processes on different qudits are independent and uncorrelated, and it is the case in many physical situations. On this ground, hereafter all explicit examples will be on local quantum channels.

\subsection{Strong vs Weak Symmetries}
\label{sec:weak-vs-strong}

At this point, it is important to clarify the distinction between two kinds of symmetries of super-operators: strong and weak symmetries. 

In the \emph{strong symmetry}, each Kraus operator $\hat{F}_{\mu}$ associated with the super-operator $\mathscr{F}$ is invariant,
\begin{math}
\mathscr{S}_\pi(\hat{F}_{\mu}) = \hat{F}_{\mu},
\end{math}
under all permutation super-operators $\mathscr{S}_\pi$.
Or, equivalently, each Kraus operator commutes with every permutation operator,
$[\hat{F}_{\mu}, \hat{\pi}]=0$.
A super-operator on two qudits of the form
\begin{equation}
\label{eq:twoQubitSuperOp}
\mathscr{F}(\hat\rho) = 
\left(\hat{f}_0\otimes\hat{f}_0\right)\hat\rho
\left(\hat{f}_0\otimes\hat{f}_0\right)^\dag
+\left(\hat{f}_1\otimes\hat{f}_1\right)\hat\rho
\left(\hat{f}_1\otimes\hat{f}_1\right)^\dag
\end{equation}
is an example.
The strong symmetry was previously exploited, e.g., in the context of quantum
communication without a shared reference frame \cite{bartlettClassicalQuantumCommunication2003}, where the selection of a reference frame corresponds to 
a specific unitary operator that is totally symmetric under exchange of any two qubits; hence, the corresponding super-operator (i.e., unitary congruence transformation) possesses the strong unitary symmetry.

On the other hand, \emph{weak symmetry} arises when the Kraus operators \(\hat{F}_{\mu}\) by themselves do not
commute with permutation operators, $[\hat{F}_{\mu},\hat{\pi}]\neq 0$,
but the super-operator as a whole remains invariant under permutations; see Eqs.~\eqref{Paper::eq:cmmSuperOp} and \eqref{Paper::eq:KrausUnitaryFreedom}.
In our context, this form of symmetry is described most conveniently in terms of the Liouville space rather than the conventional Hilbert-space formalism.
Local super-operator $\mathscr{F}$ on two-qudit system
\begin{multline}
\label{eq:localTwoQubitSuperOp}
\mathscr{F}(\hat\rho)
= 
\left(\hat{f}_0\otimes\hat{f}_0\right)\hat\rho
\left(\hat{f}_0\otimes\hat{f}_0\right)^\dag \\{}
+\left(\hat{f}_0\otimes\hat{f}_1\right)\hat\rho
\left(\hat{f}_0\otimes\hat{f}_1\right)^\dag \\{}
+\left(\hat{f}_1\otimes\hat{f}_0\right)\hat\rho
\left(\hat{f}_1\otimes\hat{f}_0\right)^\dag
\end{multline}
is an example that falls into this type.
Simple comparison of two examples~\eqref{eq:twoQubitSuperOp} and \eqref{eq:localTwoQubitSuperOp} clearly shows that the weak symmetry is far less restrictive than the strong symmetry.
For the strong symmetry, the identical Kraus operators act uniformly on all qudits, whereas for the weak, Kraus operators acting on different qudits may be different.

We remark that the weak symmetry has been previously studied in Ref.~\cite{bucaNoteSymmetryReductions2012}. However, they examined the weak symmetry under a single symmetry transformation rather than a collection of symmetry transformations forming a group.
In this work, we provide the most general formalism of the weak symmetry.

\subsection{Decoherence-Free Subsystems}
\label{sec:decoherence-free-subsystem}

Preventing decoherence as much as possible is a crucial task in manipulation and transmission of quantum information. 
In our case, where the symmetry of the quantum channel is clearly defined, the decoherence-free subsystem approach provides an efficient and generic strategy to protect the encoded quantum information against decoherence.
The particular tensor-product structure of each component in \eqref{eq:duality} allows us to take a strategy to endow only one part of the tensor product of a selected component with the decoherence-free feature and store information there.
This is far more efficient and versatile than attempting to achieve the decoherent-free feature for the whole component or for the entire density operator.
For example, a unitary operator (or generator of it) acts only on the $\mathcal{W}^\lambda$ part, keeping the $\mathcal{Y}^\lambda$ part completely intact. Therefore, any quantum information encoded in $\mathcal{Y}^\lambda$ will be protected under global unitary transformation. In this sense, one can regard $\mathcal{Y}^\lambda$ as a decoherence-free subsystem.

To achieve this goal, we use the super-Schur basis developed in Section~\ref{sec:super-schur-basis} as it is constructed based on the Schur-Weyl duality in \eqref{eq:duality}.  With the super-Schur basis at hand, it is now straightforward to identify the decoherence-free subsystems:
For a fixed partition $\lambda$ and standard Young tableau $Y^\lambda$,
the set
\begin{math}
\left\{ \kket{Y^\lambda,W^\lambda} \mid\, \forall W^\lambda \right\}
\end{math}
with all possible Weyl tableaux $W^\lambda$
forms a basis for the irreducible representation \(\mathcal{W}^{\lambda}\). 
In other words, under a permutation-symmetric super-operator \(\mathscr{F}\),
the basis elements transform as
\begin{equation}
\label{Paper::eq:10}
\mathscr{F}(\kket{Y^{\lambda},W'^{\lambda}}) 
= \sum_{W^{\lambda}}
\kket{Y^{\lambda},W^{\lambda}} A_{W^{\lambda},W'^{\lambda}}
\end{equation} 
for a fixed standard Young tableau $Y^\lambda$, where the summation is over all possible Weyl tableaux of shape $\lambda$ and
$A_{W^{\lambda},W'^{\lambda}}$ are complex numbers. 
This means that under the super-operator, a density operator \(\hat{\rho} \in \mathcal{L}(\mathcal{H}_{d}^{\otimes n})\) transforms as
\begin{equation}
\mathscr{F}(\hat{\rho}) 
= \sum_{Y^{\lambda}}\sum_{W^{\lambda},W'^{\lambda}}
\kket{Y^{\lambda},W^{\lambda}} A_{W^{\lambda},W'^{\lambda}}\;
\llangle Y^{\lambda},W'^{\lambda}\mid\hat{\rho} \rrangle,
\end{equation}
where
\begin{math}
\llangle Y^{\lambda},W'^{\lambda}\mid\hat{\rho} \rrangle	
\end{math}
is a shorthand notation for the Hilbert-Schmidt inner product [see Eq.~\eqref{Paper::eq:HSinner}] of two operators $\kket{Y^\lambda,W^\lambda}$ and $\hat\rho$.
Therefore, any permutation-symmetric super-operator is block diagonal in the super-Schur basis,
and each diagonal block features the tensor-product structure as follows
\begin{equation}
\label{Paper::eq:13}
\varF \simeq \bigoplus_\lambda \mathscr{I}^\lambda\otimes \mathscr{W}^\lambda,
\end{equation}
where $\mathscr{I}^\lambda$ is the identity super-operator acting trivially on $\mathcal{Y}^\lambda$ and $\mathscr{W}^\lambda$ a super-operator acting only on $\mathcal{W}^\lambda$.
The representation of $\varF$ in the decomposed form of Eq.~\eqref{Paper::eq:13} is a natural realization of the decompositions of the Louville space $\calL(\calH_d^{\otimes n})$ in Eq.~\eqref{eq:duality}.
It is now clear that the irreducible subspace $\mathcal{Y}^\lambda$ gives a decoherent-free subsystem that is immune to the decoherence process described by super-operator $\mathscr{F}$.

Note that the notion of decoherence-free \emph{subsystem} \cite{Zanardi00a,DeFilippo00a,Knill00a,Kempe01a,Shabani05a} is distinguished from more restrictive decoherence-free \emph{subspace} \cite{Zanardi97a,Zanardi97b,Barenco97a} in three key characteristics:
The latter usually requires a (i) \emph{strong symmetry}, and hence refers to an (ii) \emph{entire} invariant subspace of, usually, the (iii) Hilbert space $\calH_d^{\otimes n}$ (rather than the Louville space); see also Refs.~\cite{Zanardi98a,Lidar98a,Shabani05a}.
The decoherence-free subsystem, on the other hand, is allowed under a far less restrictive \emph{weak symmetry}, and exploits only the \emph{part} or \emph{factor} (e.g., $\mathcal{Y}^\lambda$) of an invariant subspace $\mathcal{Y}^\lambda\otimes\mathcal{W}^\lambda$ of the Liouville space in general as described above. 
Naturally, when the dimension of the active part $\mathcal{W}^\lambda$ is unity ($\dim\mathcal{W}^\lambda=1$), there is no distinction between the two notions.

As a matter of principle, in the context of decoherence-free subsystems, the symmetry of a quantum operation imposes a particular block-digonal form on the Kraus operators (error generators) \cite{Shabani05a}. 
However, we find it more convenient to describe, especially, the permutation symmetry of quantum operations in the form of Eqs.~\eqref{Paper::eq:cmmSuperOp} and \eqref{Paper::eq:KrausUnitaryFreedom} as described in the above.

\subsection{Three-Qubit Example}
\label{three-qubit-amplitude-damping-example}

Let us consider an example with a system of three qubits ($n=3$ and $d=2$) undergoing the amplitude damping. 
For individual qubits, the amplitude damping is described by a super-operator with single-qubit Kraus operators
\begin{equation}
\hat{f}_{0}\dot{=}
\begin{bmatrix}
1 &0\\0 & \sqrt{ 1-p }
\end{bmatrix},\quad 
\hat{f}_{1}\dot{=}
\begin{bmatrix}
0 & \sqrt{ p }\\0 & 0
\end{bmatrix},
\label{eq::16}
\end{equation}
where \(p\) represents the damping probability.
The simplest super-operator exhibiting the strong permutation symmetry is associated with Kraus operators $\hat{F}_0$, $\hat{F}_1$ and $\hat{F}_2$, where
\begin{subequations}
\label{Paper::eq:11}
\begin{align}
\label{Paper::eq:11a}
\hat{F}_0 
& := \hat{f}_{0}\otimes \hat{f}_{0}\otimes \hat{f}_{0} \\
\hat{F}_1 
& := \hat{f}_{1} \otimes \hat{f}_{1} \otimes \hat{f}_{1} ,
\end{align}
and
\begin{equation}
\label{Paper::eq:11b}
\hat{F}_2 := \sqrt{ \hat{I} 
- \hat{F}_0^\dag\hat{F}_0 
- \hat{F}_1^\dag\hat{F}_1 }
\end{equation}
\end{subequations}
In $\hat{F}_0$ and $\hat{F}_1$, all qubits experience the amplitude damping simultaneously, and
$\hat{F}_2$ is included to ensure the trace-preserving condition.
Note that Kraus operators are formally local, but physically, they are responsible for collective decoherence since all qubits are subject to the same decoherence process.
There are more general forms of super-operators with strong symmetry. Each associated multi-qubit Kraus operator is constructed by taking an arbitrary tensor product of single-qubit Kraus operators and including all its permutations in a linear combination. For example, imagine a local super-operator associated with Kraus operators
\begin{subequations}	
\begin{align}
\hat{G}_{1}
&:= \hat{f}_{1} \otimes \hat{f}_{0} \otimes \hat{f}_{0} 
+ \hat{f}_{0} \otimes \hat{f}_{1} \otimes \hat{f}_{0} 
+ \hat{f}_{0} \otimes \hat{f}_{0} \otimes \hat{f}_{1} , \\
\hat{G}_{2}
&:= \hat{f}_{1} \otimes \hat{f}_{1} \otimes \hat{f}_{0} 
+ \hat{f}_{0} \otimes \hat{f}_{1} \otimes \hat{f}_{1}
+ \hat{f}_{1} \otimes \hat{f}_{0} \otimes \hat{f}_{1} ,
\end{align}
and in addition
\begin{equation}
\hat{G}_{3} := \sqrt{ \hat{I} 
- \hat{G}_{1}^{\dagger} \hat{G}_{1}
- \hat{G}_{2}^{\dagger} \hat{G}_{2} } 
\end{equation}
to ensure the trace-preserving condition.
\end{subequations}
Unlike the example in Eq.~\eqref{Paper::eq:11}, in this example, the amplitude damping happens only on one or two qubits. However, the damping processes on different qubits are still strongly and coherently correlated.
Therefore, the quantum channel is not local as it is clear from the fact that the Kraus operators are not simple tensor products of single-qubit Kraus operators.

On the other hand, in the case of weak symmetry, there is a bigger flexibility and the amplitude dampling on individual qubits are uncorrelated (except for the constraint that the damping rates are uniform).
For example, amplitude damping acting independently on only one qubit at a time 
are described by Kraus operators of the form
\begin{multline}
\hat{f}_{0} \otimes \hat{I} \otimes \hat{I},\quad
\hat{I} \otimes \hat{f}_{0} \otimes \hat{I},\quad
\hat{I} \otimes \hat{I} \otimes \hat{f}_{0},\\
\hat{f}_{1} \otimes \hat{I} \otimes\hat{I},\quad
\hat{I} \otimes \hat{f}_{1} \otimes \hat{I},\quad
\hat{I} \otimes \hat{I} \otimes \hat{f}_{1} .
\end{multline}
In a more general scenario of weak symmetry,
mutilple qubits may be subject to amplitude damping simultaneously and yet independently, and the associated Kraus operators are given by all possible tensor products of the single-qubit Kraus operators in Eq.~\eqref{eq::16},
\begin{multline}
\hat{f}_{0} \otimes \hat{f}_{0} \otimes \hat{f}_{1},\quad
\hat{f}_{0} \otimes \hat{f}_{1} \otimes \hat{f}_{0},\quad
\hat{f}_{1} \otimes \hat{f}_{0} \otimes \hat{f}_{0}, \\
\hat{f}_{0} \otimes \hat{f}_{1} \otimes \hat{f}_{1},\quad
\hat{f}_{1} \otimes \hat{f}_{0} \otimes \hat{f}_{1},\quad
\hat{f}_{1} \otimes \hat{f}_{1} \otimes \hat{f}_{0},\\
\hat{f}_{0}\otimes \hat{f}_{0}\otimes \hat{f}_{0},\quad
\hat{f}_{1} \otimes \hat{f}_{1} \otimes \hat{f}_{1}. 
\end{multline}

Let us take a closer look at possible decoherence-free subsystems:
The total dimension of the matrix representation $\Gamma$ of a permutation-symmetric super-operator
is \((d^{2})^{3}=64\).
However, the representation is reducible, and can be decomposed into irreducible representations of smaller dimensions. Specifically, the irreducible representations correspond to integer partitions of $n=3$,
\(\lambda=\{3\}\), \(\lambda=\{2,1\}\), and \(\lambda=\{1,1,1\}\)
with dimensions 20, 20, and 4, respectively, so that the representation has the block diagonal structure
\begin{equation}
\label{Paper::eq:12}
\Gamma_{64\times 64} = 
\Gamma^{\{3\}}_{20\times 20} \oplus
\Gamma^{\{2,1\}}_{20\times 20} \oplus
\Gamma^{\{2,1\}}_{20\times 20} \oplus
\Gamma^{\{1,1,1\}}_{4\times 4} ,
\end{equation}
where for convenience we have indicated the dimensions of each block in subscripts.
It is noteworthy that there occur two identitcal irreducible blocks corresponding to \(\lambda=\{2,1\}\), and it is instructive to rewrite \eqref{Paper::eq:12} into the form
\begin{equation}
\label{Paper::eq:21}
\Gamma_{64\times 64} = 
I_{1\times 1}\otimes\Gamma^{\{3\}}_{20\times 20} \oplus
I_{2\times 2}\otimes\Gamma^{\{2,1\}}_{20\times 20} \oplus
I_{1\times 1}\otimes\Gamma^{\{1,1,1\}}_{4\times 4},
\end{equation}
in parallel with the decomposition of the Liouvill space in \eqref{eq:duality}.
It is now clear that the super-operator, i.e., $\Gamma$ acts trivially on the corresponding subsystems $\calY^\lambda$. In particular, $\calY^{\{2,1\}}$ has two dimensions, and is capable of storing quantum information in a decoherence-free manner. We refer readers to Appendix \ref{Appendix:A} for further details of the above decomposition.

\section{Symmetry in Lindblad Equation}
\label{sec:symmetry-lindblad}
 
Although the Kraus representation is the most general expression for a quantum channel, it is not practical to find Kraus operators for specific systems in realistic situations.
In many cases, the memory effects are ignorable (the Markov assumption), and then
the Lindblad equation (or quantum master equation) provides an easier way to describe and examine the decoherence dynamics of open equantum systems.
In the previous section, we investigated the symmetry properties of a quantum channel described by the Kraus representation.
In this section, for completeness, we examine the symmetry of the quantum channel in terms of the Lindblad equation. 
Since the general ideas and underlying principles are the same as for the Kraus representation, we will keep the accounts simple and summarize the formulation as a quick references.
Interestingly, a mathematical description of time-local Lindblad-type quantum master equations with permutation symmetry has been developed based on the Schur-Weyl duality, with focus on polynomial size subspaces of permutation invariant states \cite{Bastin25a}. Our work does not rely on the detailed formulation of permutation-symmetric quantum master equations.
It is also possible to examine the the decoherence-free structure of the Liouville space using the so-called peripheral subspace, a subspace of eigenoperators with egeinvalues of unit magnitude; see, e.g., Ref.~\cite{singh:2402.18703}. However, this requires computing the eigenoperators of the quantum channel, for which there is no direct symmetry-exploiting method available.

A quantum master equation is a differential equation for density operator of the form\cite{lindblad1976generators}
\begin{equation}
\label{Paper::eq:14}
\dot{\hat{\rho}} = \mathscr{L}(\hat{\rho})
\end{equation}
with the so-called Lindbladian (or simply called Lindbladian) \(\mathscr{L}\) defined by 
\begin{multline}
    \mathscr{L}(\hat{\rho}) 
:= -i \left[ \hat{H},\hat{\rho} \right]\\+\sum_{\mu}\left( \hat{L}_{\mu}
\hat{\rho}\hat{L}_{\mu}^{\dagger}-\frac{1}{2}\hat{L}_{\mu}^{\dagger}\hat{L}_{\mu}\hat{\rho}-\frac{1}{2}\hat{\rho}\hat{L}_{\mu}^{\dagger}\hat{L}_{\mu} \right),
\end{multline}
where $\hat{H}$ is the effective Hamiltonian and $\hat{L}_\mu$ are Lindblad operators (or quantum jump operators) 
in $\mathcal{L}(\mathcal{H}_d^{\otimes n})$, each labeled by an index $\mu$ and
responsible for particular decoherence processes. Throughout this work, without loss of generality, we assume that the Lindblad operators $\hat{L}_\mu$ are traceless and orthogonal to each other with
respect to the Hilbert-Schmidt product.

\subsection{Permutation symmetry in Lindblad equation}
\label{sec:perm-symmety-lindblad}

Analogous to the analysis in Section~\ref{sec:perm-symmetry-kraus}, we examine the properties of the quantum channel under permutation symmetry.
A permutation-symmetric Lindblad equation can be described by a Lindbladian that commutes with all permutation super-operators; i.e., 
\begin{math}
\mathscr{S}_{\pi}\mathscr{F}=\mathscr{F}\mathscr{S}_{\pi},
\end{math}
for all permutations $\pi$ in $\mathcal{S}_{n}$. 
In terms of the Lindblad operators, this is equivalent to requiring that
\begin{multline}
\sum_{\mu} \Bigg[ 
\mathscr{S}_{\pi}(\hat{L}_{\mu}) 
\hat{\rho} \mathscr{S}_{\pi}(\hat{L}_{\mu}^{\dagger})
- \frac{1}{2}\left\{
\mathscr{S}_{\pi}(\hat{L}_{\mu}^{\dagger} )\mathscr{S}_{\pi}(\hat{L}_{\mu}),
\hat{\rho} \right\} 
\Bigg] \\{}
= \sum_{\mu}\left[ 
\hat{L}_{\mu}\hat{\rho}\hat{L}_{\mu}^{\dagger}
- \frac{1}{2}
\left\{\hat{L}_{\mu}^{\dagger}\hat{L}_{\mu}, \hat{\rho}\right\}
\right],
\end{multline}
for all operators $\hat{\rho}$.
From the unitary freedom of the Lindblad operators~\cite{endnote:3},
this implies that for each permutation $\pi$,
there exists a unitary matrix $V(\pi)$ such that
\begin{equation}
\label{Paper::eq:symmLindbladOps}
\hat\pi\hat{L}_{\nu}\hat\pi^\dag 
= \sum_{\mu} \hat{L}_{\mu} U(\pi)_{\mu\nu}
\end{equation}
for all Lindblad operators $\hat{L}_\nu$.
Therefore, like the Kraus operators association with a permutation-symmetric super-operator [see Eq.~\eqref{Paper::eq:KrausUnitaryFreedom}],
the set of Lindblad operators $\{\hat{L}_\mu\}$ associated with the permutation-symmetric Lindbladian $\varL$ forms a representation space of the symmetric group $S_n$, and $U$ in Eq.~\eqref{Paper::eq:symmLindbladOps} is another unitary representation of $\calS_n$.
In addition, the effective Hamiltonian must be invariant under permutation,
\begin{equation}
\label{Paper::eq:symmLindbladHam}
\hat\pi\hat{H}\hat\pi^\dag = \hat{H}.
\end{equation} 

As discussed in Section~\ref{sec:weak-vs-strong}, it is instructive to distinguish two types of symmetries of Lindbladian equations as well:
the strong and weak symmetries.
In the case of \emph{strong symmetry}, every Lindblad operator $\hat{L}_{\mu}$ and the effective Hamiltonian $\hat{H}$ commute with the permutation operators $\hat{\pi}$ of a symmetric group $\mathcal{S}_{n}$ that describes the symmetry, leading to $[\hat{H},\hat{\pi}]=[\hat{L}_{\mu},\hat{\pi}]=0$. 
The weak symmetry holds when the Lindbladian as a whole remains invariant under permutations, $\mathscr{S}_{\pi}\mathscr{L}=\mathscr{L}\mathscr{S}_{\pi}$, regardless of the individual Lindblad operators $\hat{L}_{\mu}$ or the effective Hamiltonian $\hatH$.

When quantum decoherence processes are \emph{local}, the Lindblad operators are tensor products
\begin{equation}
\hat{L}_{\mu} = 
\hat\ell_{\mu_1}\otimes\cdots\hat\ell_{\mu_n}
\end{equation}
of single-qudit Lindbland operators $\ell_{\mu_k}$ of type $\mu_k$ acting on qudit $k$.
If a local quantum channel is permutation symmetric and a tensor product
\begin{math}
\hat\ell_{\mu_1}\otimes\cdots\hat\ell_{\mu_n}
\end{math}
is a Lindblad operator, 
then all its permutations
\begin{equation}
\hat\pi\left(\hat\ell_{\mu_1}\otimes\cdots\hat\ell_{\mu_n}\right)\hat\pi^\dag =
\ell_{\mu_{\pi_1}}\otimes\cdots\hat\ell_{\mu_{\pi_n}}
\end{equation}
must also be Kraus operators of the same Lindbladian in accordance with Eq.~\eqref{Paper::eq:symmLindbladOps}.
This means that a local Lindbladian may have only a weak symmetry in general~\cite{endnote:4}.

\subsection{Decoherence-Free subsystems}
\label{sec:DCS-Lindblad}

Section~\ref{sec:decoherence-free-subsystem} discusses how to systematically find the decoherence-free subsystems when the decoherence process is described in the Kraus representation.
In this subsection, we provide a similar approach to identify the decoherence-free subsystems in the framework of Lindblad equation based on symmetry.
Again, we use the super-Schur basis from Section~\ref{sec:super-schur-basis}, built on Schur-Weyl duality \eqref{eq:duality}, which simplifies the identification of decoherence-free subsystems. 

Any permutation-symmetric Lindbladian \(\mathscr{L}\) transforms the super-Schur basis elements [see also Eq.~\eqref{Paper::eq:10}] as 
\begin{equation}
\mathscr{L}(\kket{Y^{\lambda},W'^{\lambda}})
= \sum_{W^{\lambda}} \kket{Y^{\lambda},W^{\lambda}} 
B_{W^{\lambda},W'^{\lambda}}
\end{equation} 
for a fixed standard Young tableau $Y^\lambda$, where the summation is over all possible Weyl tableaux of shape $\lambda$ and
$B_{W^{\lambda},W'^{\lambda}}$ are complex numbers.
In other words, the time derivative of a density operator \(\hat{\rho} \) transforms as
\begin{equation}
\dot{\hat{\rho}}
= \sum_{Y^{\lambda}}\sum_{W^{\lambda}}\sum_{W'^{\lambda}}
\kket{Y^{\lambda},W^{\lambda}}
B_{W^{\lambda},W'^{\lambda}}\;
\aavg{Y^{\lambda},W'^{\lambda}|\hat\rho}.
\end{equation}
Furthermore, analogous to Eq.~\eqref{Paper::eq:13}, any permtutation-symmetric Lindbladian can also be represented as a block diagonal matrix in the super-Schur basis,
\begin{equation}
\label{Paper::eq:16}
\mathscr{L} \simeq \bigoplus_\lambda \mathscr{I}^\lambda\otimes \mathscr{W}^\lambda,
\end{equation}
where $\mathscr{I}^\lambda$ is the identity super-operator acting trivially on $\mathcal{Y}^\lambda$ and $\mathscr{W}^\lambda$ is a super-operator acting only on $\mathcal{W}^\lambda$.
This decomposition of the representation of \(\mathscr{L}\) is consistent with the decomposition of the Louville space \(\calL(\calH_d^{\otimes n})\) in \eqref{eq:duality}, and this structure is preserved throughout the time evolution as well, as one can see from
\begin{equation}
\begin{split}
\hat{\rho}(t) 
&= \exp (i \mathscr{L} t)(\hat{\rho}(0)) \\
&= \sum_\lambda\sum_{Y^{\lambda}}\sum_{W^{\lambda}}\sum_{W'^{\lambda}}
\kket{Y^{\lambda},W^{\lambda}} \exp(i t B)_{W^{\lambda},W'^{\lambda}} 
\\{}&\qquad\qquad\times 
\aavg{Y^{\lambda},W'^{\lambda}|\hat\rho(0)} .
\end{split}
\end{equation}
This structure of the Lindbladian $\mathscr{L}$ confirms that the irreducible subspace $\mathcal{Y}^\lambda$ constitutes a decoherence-free subsystem.

\subsection{Three-Qubit Example}
\label{three-qubit-general-damping-example }

As an example, we explore a system of three qubits undergoing the amplitude damping. The amplitude damping in individual qubits is characterized by one single-qubit Lindblad operator, 
\begin{equation} 
\sqrt{\gamma}\hat\ell \doteq \sqrt{\gamma}
\begin{bmatrix}
0 & 1 \\
0 & 0
\end{bmatrix} ,
\end{equation} where a real positive parameter $\gamma$ is the amplitude damping rate. For both strong and weak permutation symmetry, the effective Hamiltonian on the whole system should be permutation-symmetric. 
An example is the transverse-field Ising model with uniform field and coupling, which for three qubits, reads as
\begin{equation}
\hat{H} =
h_{x}\left(\hat{S}_{1}^{x}+\hat{S}_{2}^{x}+\hat{S}_{3}^{x}\right) +
J\left(\hat{S}_{1}^{z}\hat{S}^{z}_{2}+\hat{S}^{z}_{2}\hat{S}^{z}_{3}+\hat{S}^{z}_{3}\hat{S}^{z}_{1}\right),
\end{equation}
where \(h_{x}\) is responsible for an external field, $J$ is the Ising coupling between nearest neighbor qubits, and \(\hat{S}^{x,z}_{j}\) 
are the Pauli X and Z operators on qubit $j$.

As for the Kraus operators in Section~\ref{three-qubit-amplitude-damping-example}, a set of all permutations of a tensor-product Lindblad operator
\begin{equation} 
\label{Paper::eq:example-lindblad-1}
\sqrt{\gamma_{1}}\,\hat\ell \otimes \hat{I} \otimes \hat{I},\quad
\sqrt{\gamma_{1}}\,\hat{I} \otimes \hat\ell \otimes \hat{I},\quad
\sqrt{\gamma_{1}}\,\hat{I} \otimes \hat{I} \otimes \hat\ell
\end{equation}
or 
\begin{equation} 
\label{Paper::eq:example-lindblad-2}
\sqrt{\gamma_{2}}\,\hat\ell \otimes \hat\ell \otimes \hat{I},\quad
\sqrt{\gamma_{2}}\,\hat{I} \otimes \hat\ell \otimes \hat\ell,\quad
\sqrt{\gamma_{2}}\,\hat\ell \otimes \hat{I} \otimes \hat\ell
\end{equation} 
gives a weak permutation-symmetric Lindblad equations.
Here, note that the same rates \(\gamma_{1}\) and \(\gamma_{2}\) are applied to every operator identically. 

For a comparison, let us construct Lindbland operators associated with a \emph{strong} permutation-symmetric Lindblandian. Since each Lindblad operator must be invariant under all permutations,
possible examples include
\begin{multline}
\sqrt{\gamma_{3}}
\left(\hat\ell \otimes \hat{I} \otimes \hat{I}+ \hat{I} \otimes \hat\ell \otimes \hat{I}+ \hat{I} \otimes \hat{I} \otimes \hat\ell \right) , \\
\sqrt{\gamma_{4}}
\left(\hat\ell \otimes \hat\ell \otimes \hat{I}+ \hat{I} \otimes \hat\ell \otimes \hat\ell+\hat\ell \otimes \hat{I} \otimes \hat\ell\right)
\label{Paper::eq:example-lindblad-3}
\end{multline}
and
\begin{equation}
\label{Paper::eq:example-lindblad-4}
\sqrt{\gamma_{5}}\,\hat\ell \otimes \hat\ell \otimes \hat\ell,
\end{equation}
where $\gamma_3$, $\gamma_4$ and $\gamma_5$ are the rates of corresponding decoherence processes.
Unlike the examples~\eqref{Paper::eq:example-lindblad-1} and \eqref{Paper::eq:example-lindblad-2},
the Lindblad operators in Eq.~\eqref{Paper::eq:example-lindblad-3} is not a tensor product.
The Lindblad operator in Eq.~\eqref{Paper::eq:example-lindblad-4} is formally local, but all qubits experience an identical decoherence process and the docoherence process is collective. Therefore, these examles illustrate that a noisy quantum channel with strong symmetry is not local.

All these considerations in the form of allowed Lindblad operators lead to the block-diagonal structure of the $64 \times 64$ matrix representation $\Gamma$ of the Lindbladian $\mathscr{L}$ (see Appendix~\ref{Appendix:A} for details)
in the same way as for the super-operator $\varF$ discussed in Section~\ref{three-qubit-amplitude-damping-example},
\begin{equation}
\Gamma_{64\times 64} = 
I_{1\times 1}\otimes\Gamma^{\{3\}}_{20\times 20} \oplus
I_{2\times 2}\otimes\Gamma^{\{2,1\}}_{20\times 20} \oplus
I_{1\times 1}\otimes\Gamma^{\{1,1,1\}}_{4\times 4}.
\end{equation}
This structure of the matrix representation is preserved in the matrix exponential as well,
\begin{multline}
\exp(i t \Gamma_{64\times 64}) = 
I_{1\times 1}\;\otimes\; \exp\left(i t\Gamma^{\{3\}}_{20\times 20}\right) \\{} \oplus
I_{2\times 2}\;\otimes\; \exp\left(i t\Gamma^{\{2,1\}}_{20\times 20}\right)\\{} \oplus
I_{1\times 1}\;\otimes\; \exp\left(i t \Gamma^{\{1,1,1\}}_{4\times 4}\right).
\end{multline}
This clearly illustrates that the subsystem $\calY^{\{2,1\}}$ is a decoherence-free subsystem.

\vspace{\baselineskip}

\section{Conclusion}
\label{conclusion}

We have proposed a generic scheme to encode quantum information into an open quantum system based on symmetry.
Under a given symmetry, the Liouville space is decomposed into invariant subspaces featuring a tensor-product structure.
A decoherence-free subsystem is then identified with a factor of the tensor product.
Unlike decoherence-free subspaces, which require strong symmetries, decoherence-free systems are allowed in far less restrictive weak symmetries.
To be specific, we primarily focused on the permutation symmetry in conjunction with the unitary symmetry, and utilized the Schur-Weyl duality,
which facilitates many efficient and systematic group-representation theoretic calculations.
Using the isomorphism between the Liouville space $\mathcal{L}(\mathcal{H}_d^{\otimes n})$ and the ficticious Hilbert space $\mathcal{H}_{d^2}^{\otimes n}$, we constructed a super-Schur basis, which block-diagonalizes the super-operators that describe the noisy quantum channels, either in the Kraus representation or in terms of quantum master equation.
Each block reveals the tensor-product structure and allows us to easily identify physically relevant decoherence-free subsystems under the given weak symmetry.

\section*{Acknowledgments}

This work has been supported by the National Research Function (NRF) of Korea (Grant Nos. 2022M3H3A1063074 and 2023R1A2C1005588) and by the BK21 Program.

\onecolumngrid
\appendix

\section{Block Diagonalization of the Matrix Representations of Permutation Symmetric Super-Operators}
\label{Appendix:A}

In this appendix, we demonstrate the structure of the permutation symmetric quantum channels through matrix representations of their corresponding super-operators, which we refer to as \emph{super-matrices} for convenience. We focus on examples from the 3-qubit system discussed in Sec.~\ref{sec:symmetry-kraus} and \ref{sec:symmetry-lindblad}. Since each qubit has dimension $d=2$, the total dimension of the super-matrix is $64 \times 64$ from the dimension of the Liouville space $2^{2\times(n=3)}=64$.

\begin{figure}[htbp]
\centering
\includegraphics[width=180mm]{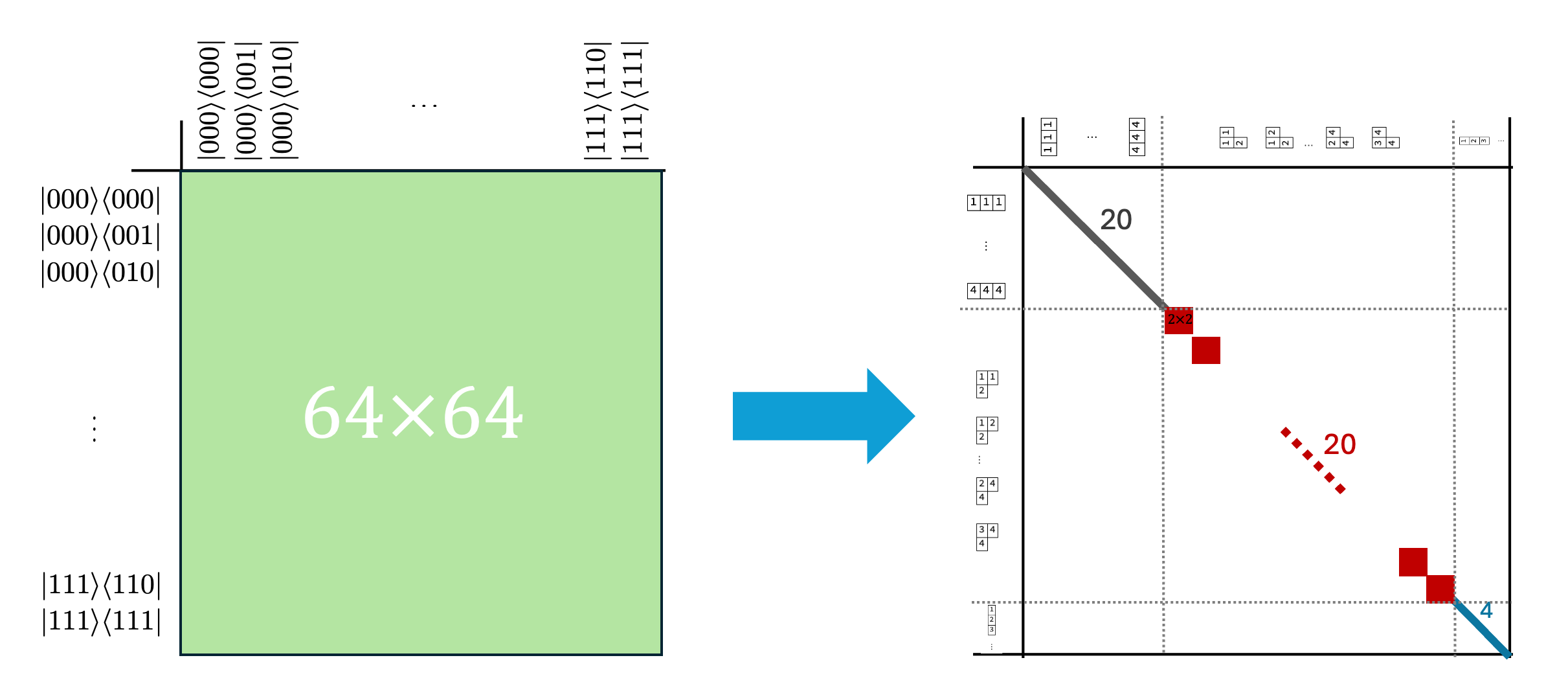}
\caption{Block diagonalization of the matrix representations of permutation-symmetric super-operators. The square on the left depicts the \(64 \times 64\) super-matrix for the symmetric group \(\mathcal{S}_3\) in the computational basis. The blue arrow indicates the transformation to the super-Schur basis, sorted by the Weyl tableaux as displayed along the sides of the square on the right. 
The square on the right contains structured blocks: black and blue lines denote one-dimensional blocks, corresponding to simple diagonalization, while red filled boxes represent two-dimensional blocks, indicating nontrivial block diagonalization. These blocks correspond to different irreducible representations, with gray, red, and blue colors denoting distinct representations. The numbers indicate the multiplicities of each representation, determined by the number of Weyl tableaux.}
\label{fig:perm_block_diag}
\end{figure}

We first examine the permutation super-operators to understand the permutation symmetry in the super-matrices. The super-Schur basis in Sec. \ref{sec:super-schur-basis} is instrumental in achieving the block diagonalization of these super-matrices, as illustrated in Fig.~\ref{fig:perm_block_diag}.

Both the left and right squares in Fig.~\ref{fig:perm_block_diag} summarize the super-matrices corresponding to a permutation super-operator that represents the symmetric group $\mathcal{S}_{3}$. The operators on the two sides of the left square represent the computational basis. On this basis, the super-matrices generally do not exhibit block diagonalization, a feature demonstrated by the large green square labeled `$64\times 64$', which indicates the total dimension of the super-matrix. 

The blue arrow in Fig.~\ref{fig:perm_block_diag} illustrates the transformation from the computational basis to the super-Schur basis. The super-Schur basis is sorted by the order of the Weyl tableaux, as shown on the two sides of the right square. This transformation block diagonalizes every permutation super-matrix.  

Three distinct irreducible representations appear by the Schur-Weyl duality in Eq.\eqref{eq:duality}. The number of distinct irreducible representations is identical to the number of partitions in Eq.\eqref{eq:duality}; see Appendix~\ref{Appendix:B} for details. Those representations are denoted as gray, red, and blue in the right square. The dimension of each representation is identical to the number of standard Young tableaux. The gray and blue representations correspond to the partitions $\{3\}$ and $\{1,1,1\}$ respectively, that each representation is one-dimension.
The red representation, corresponding to the partition $\{2,1\}$, is two-dimensional; these are represented as small squares aligned along the big square's diagonal. The numbers adjacent to each color indicate the frequency of each representation, equivalent to the number of Weyl tableaux. 

\begin{figure}[bt]
    \centering
    \includegraphics[width=180mm]{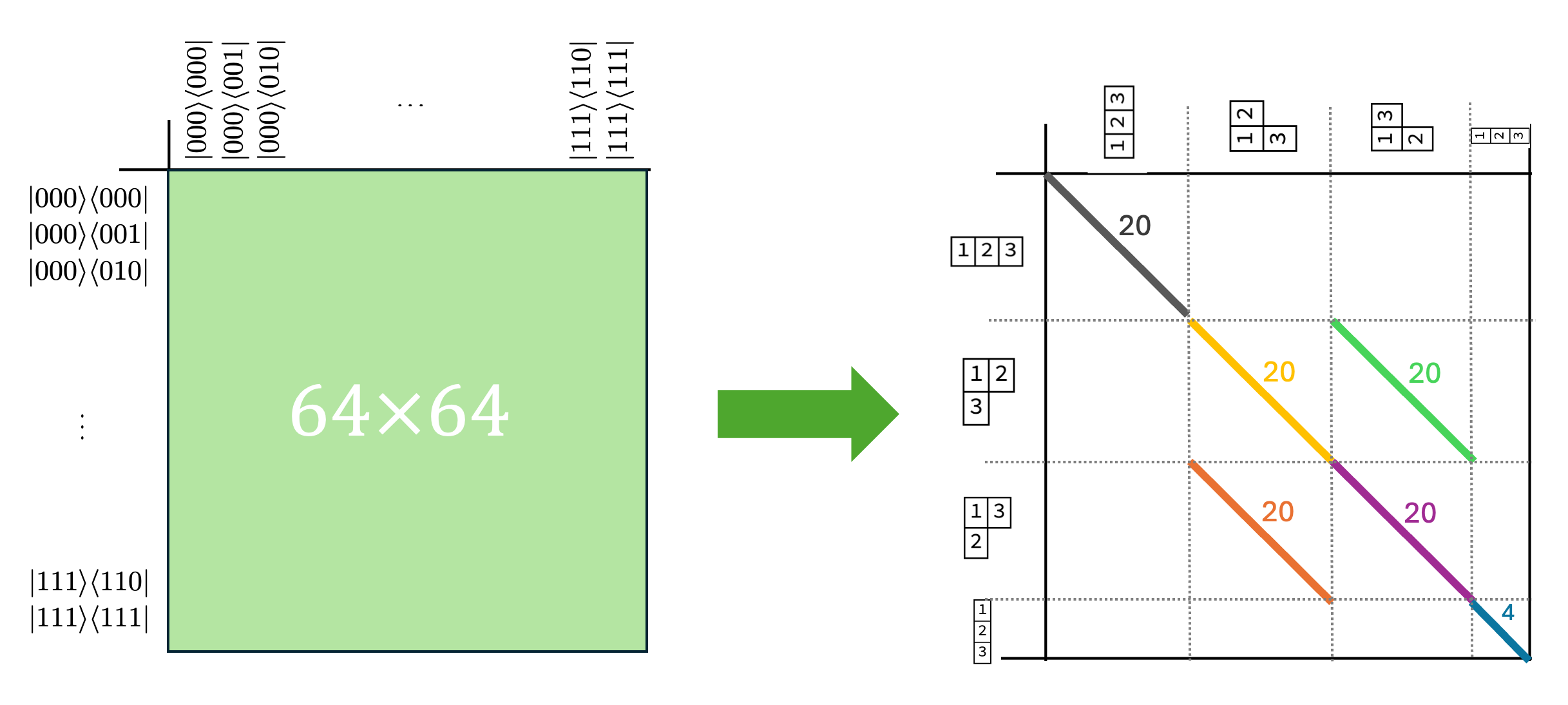}
    \caption{Block diagonalization of permutation super-matrix with distinct structure from Fig.~\ref{fig:perm_block_diag}. The left square, identical to Fig.~\ref{fig:perm_block_diag}, shows the \(64 \times 64\) super-matrix for \(\mathcal{S}_3\) in the computational basis. The green arrow indicates the transformation to the super-Schur basis, sorted by standard Young tableaux, displayed in the right square. 
    The result of the transformation is also block diagonalized by each Young diagram.
     The one-dimensional representations (gray and blue lines) remain the same, while the two-dimensional representations in Fig.~\ref{fig:perm_block_diag} are reorganized into distinct blocks, distinguished by  Each diagonal line within the red dashed blocks representing a scalar multiple of the identity matrix, with numbers indicating their dimensions.}
    \label{fig:perm_not_block}
\end{figure}

However, we can also consider a basis transformation where the super-Schur basis is sorted by the Standard Young Tableaux instead. This transformation is illustrated by the green arrow in Fig.~\ref{fig:perm_not_block}. 
        
In Fig.~\ref{fig:perm_not_block}, the rearranged super-Schur basis is represented on the two sides of the right square.
The blocks, marked with gray dashed lines, once again demonstrate the block diagonalization of the permutation super-matrix. This structure is determined by the partitions, as indicated by the standard Young Tableaux on both sides of the right square.

The one-dimensional representations in Fig.~\ref{fig:perm_block_diag}, depicted as gray and blue lines, are identical in Fig.~\ref{fig:perm_not_block}. However, the two-dimensional representations shown as small red squares in 
Fig.~\ref{fig:perm_block_diag} are rearranged diagonal lines in Fig.~\ref{fig:perm_not_block}. Each diagonal line represents a scalar multiple of the identity matrix.
The numbers adjacent to each line indicate the dimensions of each diagonal, which is identical to the number of Weyl tableaux.
Therefore, this super-matrix contains only scalar multiples of the identity matrices. The configuration of the permutation super-matrices makes it easier to understand their commutation properties, leading us to examine the permutation symmetric super-matrix next.

\begin{figure}[bt]
    \centering
    \includegraphics[width=180mm]{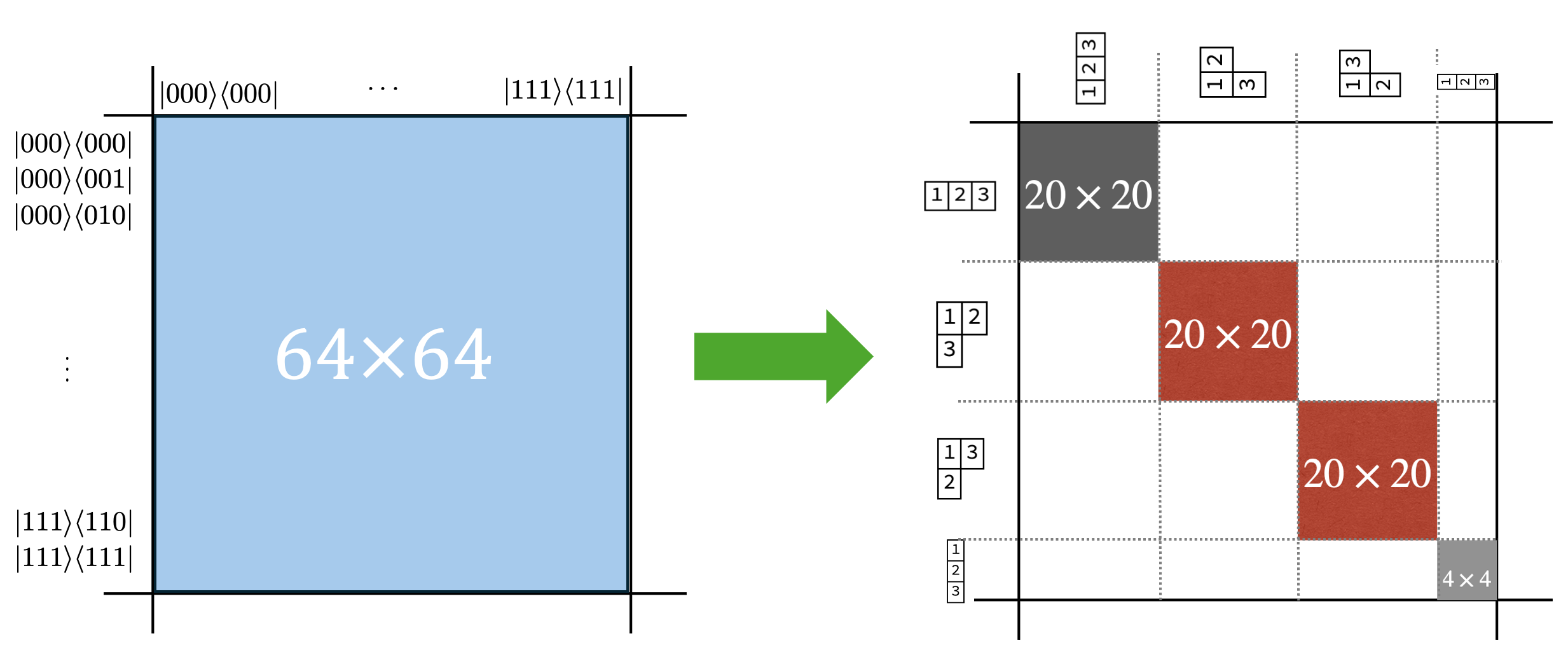}
    \caption{Block diagonalization of permutation symmetric super-matrix. The left blue square represents the \(64 \times 64\) permutation symmetric super-matrix in the computational basis, which generally lacks block diagonalization. The green arrow, identical to Fig.~\ref{fig:perm_not_block}, indicates the transformation to the super-Schur basis sorted by standard Young tableaux, as shown in the right square. The right square displays four blocks demonstrating block diagonalization. Each block is labeled as "20 × 20" or "4 × 4" based on size. Gray blocks align with dashed lines in Fig.~\ref{fig:perm_not_block}, ensuring commutation with permutation super-matrices. The two middle red blocks must be identical to maintain this commutation.}
    \label{fig:perm_inv_diag}
\end{figure}

We can apply the same basis transformation as shown in Fig.~\ref{fig:perm_not_block} to the matrix representation of the permutation symmetric super-operators discussed in Sec.~\ref{sec:symmetry-kraus} and \ref{sec:symmetry-lindblad}. 
The basis transformation represented by the green arrow in Fig.~\ref{fig:perm_not_block} is also depicted in Fig.~\ref{fig:perm_inv_diag}, illustrating the identical basis transformation for the permutation symmetric super-matrix.

In Fig.~\ref{fig:perm_inv_diag}, the blue square on the left represents the super-matrix in the computational basis, where the super-matrices generally do not exhibit block diagonalization. Four blocks in the right square demonstrate the block diagonalization on a super-Schur basis. Each block is labeled with either ``$20 \times 20$" or ``$4 \times 4$" depending on the size of the block diagonal matrix, which is identical to the number of Weyl tableaux.
The gray blocks are occupied at the same position as those with dashed lines in Fig.~\ref{fig:perm_not_block}, ensuring that these blocks always commute with the permutation super-matrices regardless of their entities. The middle two blocks, both colored red, indicate that they should be identical to ensure commutation with the permutation super-matrices. This feature can also be derived from the Schur-Weyl duality formulated in Eq.\eqref{eq:duality}. 

From the identical block diagonal structure in Fig.~\ref{fig:perm_inv_diag}, we can directly identify the decoherence-free subsystem within any permutation symmetric Kraus representation. Consider a vector on a super-Schur basis corresponding to the input density operator of the quantum channel associated with a permutation symmetric Kraus representation. In the vector, encode information between two standard Young tableaux of the partition $\left\{ 2,1 \right\}$, which undergoes the same decoherence process. This uniformity ensures that the fidelity of the information remains unchanged, leading to a decoherence-free subsystem. Therefore, if the permutation symmetry of the Kraus representation is confirmed, you can directly identify the decoherence-free subsystem with the super-Schur basis without examining the specific details of the Kraus operators.

The block diagonal structure shown in Fig.~\ref{fig:perm_inv_diag} is also consistent with the permutation-symmetric Lindblad equation since the time evolution is expressed as the exponential function of the super-matrix. 

\section{The Number of Irreducible Representations}
\label{Appendix:B}

In this section, we demonstrate the systematic way to calculate the number of 
partitions in Eq.\eqref{eq:duality}, which corresponds to the number of irreducible representations in Fig.~\ref{fig:perm_block_diag}.

From the Sec. \ref{schur-weyl-duality-in-liouville-space}, the total number of inequivalent irreducible Liouville subspaces is determined by the
the number of partitions $\lambda$ of an \(n\) with $d^2$ rows.

To systematically count all allowed partitions (i.e., irreducible Liouville subspaces), we introduce a generating function from \cite{hardy1979introduction}.

Specifically, let \(p_{k}(n)\) denote the number of Young diagrams with
\(k\) rows (\(k \leq n\)). The generating function for \(p_{k}(n)\) is expressed as
\begin{equation}
\sum\limits_{n\geq0}p_{k}(n)x^{n}=x^{k}\prod\limits_{i=1}^{k} \frac{1}{1-x^{i}}.
\end{equation}
Then, the number of components in Eq.\eqref{eq:duality}, or equivalently, the total number of partitions of $n$ into at most $d^2$ integers,
can be obtained as
\begin{equation}
\sum_{k=1}^{\min[n,d^{2}]}p_{k}(n).
\end{equation} 
For instance, in a 3-qubit system, 
the above generating function gives
\(p_{1}(3)=p_{2}(3)=p_{3}(3)=1\). Therefore, the number of all possible
\(\lambda\) is 3 in this case.

\twocolumngrid
\bibliographystyle{apsrev}
\bibliography{References}

\begin{thebibliography}{49}
\bibitem[{end({\natexlab{a}})}]{endnote:ChoiIsomorphism}
\bibinfo{note}{The correspondence $\mathcal{L}(\mathcal{H}_d^{\otimes
  n})\simeq\mathcal{H}_{d^2}^{\otimes n}$ may be rigorously described by
  utilizing the Choi isomorphism (also known as Jamiolkowski-Choi or
  Choi-Jamiolkowski isomorphim), where a density opeerators correspond to a
  vector and super-operators are mapped to usual operators
  \cite{Nielsen11a,Choi75a,Jamiolkowski72a}.}
\bibitem[{end({\natexlab{b}})}]{endnote:1}
\bibinfo{note}{Recall that the system is assumed to have only the permutation
  symmetry but not the unitary symmetry. However, the unitary symmetry is also
  considered here for technical convenience as per the Schur-Weyl duality.}
\bibitem[{end({\natexlab{c}})}]{endnote:2}
\bibinfo{note}{Two super-operators $\mathscr{F}$ and $\mathscr{G}$ associated
  with Kraus operators $\hat{F}_\mu$ and $\hat{G}_\nu$, respectively, are
  identical, i.e., \begin{math} \sum_{\mu}\hat{F}_{\mu} \hat{\rho}
  \hat{F}_{\mu}^{\dagger} =\sum_{\nu}\hat{G}_{\nu}\hat{\rho}
  \hat{G}_{\nu}^{\dagger} \end{math} as long as
  \(\hat{G}_{\nu}=\sum_{\mu}\hat{F}_{\mu}U_{\mu \nu}\) for some unitary matrix
  \(U\). See Refs.~\cite{Nielsen11a,ChoiMS22a} for more details.}
\bibitem[{end({\natexlab{d}})}]{endnote:3}
\bibinfo{note}{Two Lindbland super-operators $\varL$ and $\varM$ associated
  with Lindblad operators $\hat{L}_\mu$ and $\hat{M}_\nu$, respectively, are
  identical if \begin{math} \hat{M}_{\nu} = \sum_{\mu} \hat{L}_{\mu} U_{\mu
  \nu} \end{math} for some unitary matrix $U$. In fact, the Lindblad equation
  is also invariant under the inhomogeneous transformations of Lindblad
  operators \begin{math} \hat{L}_{\mu} \to \hat{L}_{\mu}+a_{\mu} \end{math} and
  Hamiltonian \begin{math} \hat{H} \to \hat{H} + \frac{1}{2i} \sum_{\mu}
  \left(a^{*}_{\mu} \hat{L}_{\mu} -a_{\mu} \hat{L}_{\mu}^{\dagger}\right) + b
  \end{math} where $a_\mu$ are complex numbers and $b$ is a real nubmer.
  However, this translational freedom is irrelevant for symmetries such as the
  permutation symmetry of our interest. See \cite{Breuer02a,ChoiMS22a} for more
  details.}
\bibitem[{end({\natexlab{e}})}]{endnote:4}
\bibinfo{note}{Lindblad operators with identical factors, \begin{math}
  \hat\ell_{\mu_k}\otimes\cdots\otimes\hat\ell_{\mu_k} \end{math} is formally
  local. Howerver, physically such a form is responsible for collective
  coherence with all qudits are subject to the same decoherence process, and
  the quantum channel is not local in physical nature.}
\end{thebibliography}

\end{document}